# Extended Experimental Inferential Structure Determination Method for Evaluating the Structural Ensembles of Disordered Protein States

James Lincoff[1,2,+], Mickael Krzeminski[3], Mojtaba Haghighatlari[2,6], João M.C. Teixeira[3,4], Gregory-Neal W. Gomes[5], Claudiu C. Gradinaru[5], Julie D. Forman-Kay[3,4], Teresa Head-Gordon[1,2,6*]

[1]Department of Chemical and Biomolecular Engineering and [2]Pitzer Center for Theoretical Chemistry, University of California, Berkeley, CA 94720
[3]Molecular Structure and Function Program, Hospital for Sick Children, Toronto, Ontario M5G 0A4, Canada
[4]Department of Biochemistry, University of Toronto, Toronto, Ontario M5S 1A8, Canada
[5]Department of Chemical and Physical Sciences, University of Toronto, Mississauga, Mississauga, Ontario L5L 1C6, Canada
[6]Department of Chemistry and [7]Department of Bioengineering, University of California, Berkeley, CA 94720

Characterization of proteins with intrinsic or unfolded state disorder comprises a new frontier in structural biology, requiring the characterization of diverse and dynamic structural ensembles. We introduce a comprehensive Bayesian framework, the Extended Experimental Inferential Structure Determination (X-EISD) method, that calculates the maximum log-likelihood of a protein structural ensemble by accounting for the uncertainties of a wide range of experimental data and back-calculation models from structures, including NMR chemical shifts, J-couplings, Nuclear Overhauser Effects, paramagnetic relaxation enhancements, residual dipolar couplings, and hydrodynamic radii, single molecule fluorescence Förster resonance energy transfer efficiencies and small angle X-ray scattering intensity curves. We apply X-EISD to the drkN SH3 unfolded state domain and show that certain experimental data types are more influential than others for both eliminating structural ensemble models, while also finding equally probable disordered ensembles that have alternative structural properties that will stimulate further experiments to discriminate between them.

[+]Current address: Cardiovascular Research Institute, University of California, San Francisco, CA 94158
*corresponding author: thg@berkeley.edu

**INTRODUCTION**

Experimental techniques such as X-ray and electron crystallography and microscopy, which have traditionally excelled at determining the atomic structures of protein macromolecules and their complexes, are ill-suited for analysis of proteins with intrinsic or unfolded state disorder.[1] Instead the degree to which a simulated conformational ensemble for an intrinsically disordered protein (IDP) or unfolded state of a protein can be trusted to represent functionally relevant conformations is judged by the extent to which it conforms to the information available from solution experimental data.[1, 2] But generating and validating structural ensembles for IDPs and unfolded ensembles has proven challenging due to many factors.[3] First is the need for multiple experimental data types that probe both local and global disorder, necessary given the under-determined nature of experiments that can only measure time and/or ensemble averages.[4] Since the differential value of each experimental data type for refining computational ensembles is not well established, and the fact that the accuracy of a back-calculation from the set of simulated conformers to an observable adds additional uncertainty, the quality of a constructed disordered ensemble is not necessarily captured through standard evaluation metrics such as mean squared errors, correlation coefficients, or other figures of merit. While a number of Bayesian statistical models have been put forth to determine the most probable structural ensemble for ordered to disordered states[5-7], they do not fully account for different sources of uncertainty that varies by type of experiment and the back-calculation model used in the validation process.

We introduce a complete Bayesian model, the extended Experimental Inferential Structure Determination (X-EISD) method, for the statistical modeling of a wide range of experimental data types for proteins with disordered states: NMR chemical shifts and J-couplings[8], homonuclear Nuclear Overhauser effects (NOEs)[3, 9, 10], paramagnetic relaxation enhancements (PREs)[11, 12], residual dipolar couplings (RDCs)[13, 14], hydrodynamic radii ($R_h$)[15], transfer efficiencies from single-molecule Förster resonance energy transfer (smFRET) experiments[16, 17], and small-angle X-ray scattering (SAXS) intensity curves[18, 19]. We apply the X-EISD procedure on the unfolded state of the drkN SH3 domain because of the wide variety of experimental data types made available by the Forman-Kay and Gradinaru groups[15, 20], and which has made it popular as a test system for other ensemble scoring and refinement programs[21]. Starting from either an unoptimized random coil ensemble as well as a reported structural ensemble of the unfolded state of the drkN SH3 domain[22], we show through a series of cross-validation tests the relative influence of the different data types in scoring the putative structural ensembles. With further refinement using a straightforward Markov Chain Monte Carlo (MCMC) procedure on a mixed ensemble on a spectrum of disordered to ordered conformations, we show that the extensive experimental data set supports two equally probable ensembles, but each yielding an alternative structural view that can stimulate further experiments. The X-EISD Bayesian method can be downloaded and run stand alone from a publicly available github repository (https://thglab.berkeley.edu/software-and-data/) or as part of the ENSEMBLE program[23].

**RESULTS**

The X-EISD method is formulated as a generalized Bayesian Model

$$log\ p(X,\xi|D,I) = log\ p(X|I) + \sum_{j=1}^{M} log[p(d_j|X,\xi_j,I)p(\xi_j|I)] \qquad (1)$$

where $log\ p(X,\xi|D,I)$ is the log likelihood that the ensemble of $N$ conformations $X = \{x_i\}_{i=1}^{N}$ are in agreement with the set of $M$ experimental values $D = \{d_j\}_{j=1}^{M}$, given back-calculation error and experimental uncertainties $\{\xi\}$, and any related prior information $I$. The structural prior $p(X|I)$ can be treated as either an uninformative prior or a structural prior based on Boltzmann weighting; in this work we use Jeffries uninformative prior as discussed previously[8] unlike other Bayesian methods which assumes a Boltzmann weighted ensemble[7]. It is important to state that the prior distribution $p(\xi_j|I)$ represents the uncertainty for *each* experimental and/or back-calculation nuisance parameter $\xi_j$ for data point $j$; to reflect the variable uncertainties for each data type, the nuisance parameters are treated as a Gaussian random variable as described previously.[8] Finally, $p(d_j|X,\xi_j,I)$ models the experimental data point $d_j$ given a set of conformers and model for $\xi_j$ for each data point $j$. Applying the maximum likelihood estimator, the total probability is the sum over all data points. A prototype EISD method was previously developed utilizing only J-coupling (JC) and chemical shift (CS) data for both folded proteins and IDPs[8]. These two data types illustrate two general ways to formulate the probabilistic uncertainties for any experimental observable that utilizes different types of model for the back-calculation, which is often not taken into account in other Bayesian methods.

*J-Couplings.* The Karplus equation[24, 25] is used to back-calculate the $J$ scalar coupling

$$J = A(cos(<\phi>-\phi_o))^2 + B\ cos(<\phi>-\phi_o) + C \qquad (2)$$

in which the $N$ conformations provide an ensemble-averaged dihedral angle $<\phi>$ with respect to a reference state $\phi_o$, and Eq. (2) is used to compare to the experimentally determined value. In this case the $A(\mu_A, \sigma_A)$, $B(\mu_B, \sigma_B)$, and $C(\mu_C, \sigma_C)$ are back-calculation $\xi_j$ parameters treated as Gaussian random variables for which the mean values $\mu_j$ and standard deviation $\sigma_j$ are provided in the work of Vuister and Bax.[26] The deviation of the back-calculated $J$ from the given experimental $J$ value, $\epsilon_{ex}^J$

$$\epsilon_{ex}^J(0, \sigma_J) = D_J - (A(\cos(<\phi> - \phi_o))^2 + B\cos(<\phi> - \phi_o) + C) \quad (3)$$

is also treated as a Gaussian random variable drawn from a distribution with mean 0 and a standard deviation $\sigma_{Jex}$ that estimates the experimental uncertainty of the $J$ measurement; in this work $\sigma_{Jex}=0.5$ based on the J coupling data for the drkN SH3 domain.[15] Hence the X-EISD method optimizes over all four sources of uncertainty

$$\log p(J|I) = \log p(A|\mu_A, \sigma_A) + \log p(B|\mu_B, \sigma_B) + \log p(C|\mu_C, \sigma_C) + \log p(\epsilon_{ex}^J|0, \sigma_{Jex}) \quad (4)$$

*Chemical Shifts.* The approach for chemical shifts is different, because the common back-calculators, such as SHIFTX2[27] and SPARTA+[28], incorporate their own internal weighting for the different components used to back-calculate chemical shifts, $\delta$, for each atom type, $\alpha$, that precludes a simple mathematical form such as the Karplus equation. For this reason the chemical shift back-calculator is treated as a black-box model that optimizes over $q_{\delta_\alpha}$ which is treated as a Gaussian random variable with mean 0 and standard deviation $\sigma_{q_{\delta_\alpha}}$; we use the published root-mean-square deviation (RMSD) for SHIFTX2[27] which varies from $\sigma_{q_{\delta_\alpha}}=0.1$-$0.5$ ppm depending on the relevant atom types. The chemical shift function $\epsilon_{ex}^{\delta_\alpha}$

$$\epsilon_{ex}^{\delta_\alpha}(0, \sigma_{\delta_\alpha ex}) = D_{\delta_\alpha} - q_{\delta_\alpha} - <\delta_\alpha> \quad (5)$$

is the difference between the experimental chemical shift value $D_{\delta_\alpha}$ and the average of the back-calculated shifts $<\delta_\alpha>$ from each structure of the ensemble, and accounting for the back-calculation error $q_{\delta_\alpha}$; in this work it is also treated as a Gaussian random variable drawn from a distribution with mean 0 and standard deviation $\sigma_{\delta_\alpha ex}$ that represents the experimental uncertainty of the chemical shift measurement; we assume a standard value of $\sigma_{\delta_\alpha ex}= 0.3$ ppm for C, C$_\alpha$, and C$_\beta$ and 0.03 ppm for H and H$_\alpha$. Hence the X-EISD method for chemical shifts optimizes over

$$log\, p(\delta_\alpha|I) = log\, p\left(q_{\delta_\alpha}|0, \sigma_{q_{\delta_\alpha}}\right) + log\, p\left(\epsilon_{ex}^{\delta_\alpha}|0, \sigma_{\delta_\alpha ex}\right) \qquad (6)$$

*Nuclear Overhauser Effects (NOEs).* Characterization of NOEs for IDPs is more complex than for folded proteins due to the decreased ability to precisely assign peak values to specific nuclei due to structural ensemble averaging effects[29]. Furthermore, back-calculation of NOEs from simulation can be done to varying degrees of rigor, depending on whether or not dynamical information is available and incorporated.[3] When the conformational ensemble is derived from molecular dynamics, it is possible to fully incorporate the dynamical effects on NOEs as we have shown previously.[3, 9, 10] These in turn are used to calculate per-conformer estimates of the spectral density functions, allowing fairly precise back-calculation of, for example, homonuclear $^1$H-$^1$H and heteronuclear $^1$H-$^{15}$N NOEs, and R1 and R2 relaxation times.[30] When using only static structures generated with statistical coil models such as TraDES[31] or Flexible-Meccano[32], or any other technique where no dynamical information is available, direct back-calculation is less rigorous. In this case homonuclear NOEs can be interpreted as providing information on the distance between two spins[3, 5, 9], such as the hydrogen-hydrogen distance for homonuclear $^1$H-$^1$H NOEs to estimate the scaled, ensemble-averaged values of the peak intensity.

Most standard NMR spectroscopy analysis packages[33-35] convert NOE intensities to distance restraints of varying tightness between a single pair of atoms, or pairs of atoms if the peak assignment is ambiguous. In many cases distance restraints are further binned into classes, such as strong restraints of < 2.5 Å, medium restraints < 4 Å, and weak restraints < 5 Å. Given this common classification into distance classes, the X-EISD method adopts the same approach to back-calculation as ENSEMBLE[4, 21-23], calculating the ensemble-averaged distance $\langle D \rangle$ from the set of $N$ structures

$$\langle D \rangle = \left\langle \left(\frac{\sum_{i=1}^{N} d_i^{-6}}{N}\right)^{-1/6} \right\rangle \qquad (7)$$

and the deviation between experimental and back-calculation $\epsilon_{ex}$ is calculated as

$$\epsilon_{ex}^{NOE}(0, \sigma_{NOEex}) = D_{NOE} - q_{NOE} - \langle D \rangle \qquad (8)$$

in which $q_{NOE}$ and $\epsilon_{ex}^{NOE}$ are Gaussian random variables, with mean 0 and standard deviations $\sigma_{qNOE}$ and $\sigma_{NOEex}$, similar to that used for chemical shifts. Hence X-EISD optimizes over

$$log\, p(D_{NOE}|I) = log\, p\left(q_{NOE}|0, \sigma_{qNOE}\right) + log\, p(\epsilon_{ex}^{NOE}|0, \sigma_{NOEex}) \qquad (9)$$

for every distant restraint. In order to assign the target value $D_{NOE}$ derived from a NOE measurement, we define it to be the midpoint of the experimental distance restraint, in this case either 4 or 5 Å given that the reported NOEs for the drkN SH3 domain have upper bounds of 8 or 10 Å. Note that these data were derived from largely deuterated samples using long NOE mixing times, in order to increase the likelihood of NOEs representing contacts between residues far apart in sequence, and leading to longer distance restraints than typical for standard folded protein NOEs.[36, 37] Because our simple back-calculation is effectively just a comparison of ensemble-averaged simulation distances to processed experimental distance restraints, we set the back-calculation to a small value of $\sigma_{qNOE} = 0.001$ Å. To define $\sigma_{NOEex}$ we have tested multiple uncertainty estimates based on dividing the distant restraint range by a series of integers. The resulting relative probabilities of an observed distance, normalized to the restraint value, are shown in Figure S1. Ultimately we have found that the X-EISD optimized outcome is not particularly sensitive to the value $\sigma_{NOEex}$ and have used the looser value of 4 or 5 Å.

*Paramagnetic relaxation enhancements (PREs).* Similar to NOEs, PREs report on ensemble- and time-averaged distances with strong dynamical contributions, but unlike NOEs the PRE signals can be measured for a much larger range of distances, 10 – 25 Å. [13, 38]. To conduct PRE experiments, a paramagnetic center must be introduced to the protein, such as through covalent bonding of a spin label, commonly MTSL for IDPs. The experiment then reports differences in the relaxation rates between the paramagnetic active sample versus its diamagnetic analogue, which are converted to estimates of distances between the paramagnetic center and, most commonly, the amide protons of each residue. Multiple constructs with the tag at different locations on the protein may be used to provide several sets of restraints. As with NOEs, PREs are often converted to generic long distance restraints over a range of 25 – 100 Å, to short distance restraints less than 10 Å, and a set of medium-range distance restraints, where the signal response is strongest with respect to distance, 10 – 25 Å [39]. One potential issue with PREs is whether the chemical modification of system induces different dynamics, or alters the weighting and/or introduces new structural sub-populations in the IDP ensemble[12]; at the same time, careful selection of the tag and its location can be used to minimize this potential for experimental error. Hence we assume the same X-EISD model for PREs as for NOEs, with $\sigma_{qPRE} = 0.001$ Å, but using $\sigma_{PREex}$ that divides the experimentally-derived restraint range by 4 to fit the range to a 95 % confidence interval on the normal distribution.

*Residual Dipolar Couplings (RDCs)*. Dipolar couplings between pairs of spins can provide useful signals for predicting local structure by inducing partial alignment of molecules in solution with magnetic field[13, 14]. For IDPs, RDCs resulting from the alignment of the amide in the peptide bond are the most commonly measured and reported. Back-calculation of RDCs uses either a global alignment tensor of the static structures for the entire protein as in PALES[40], or locally using fragments of the protein as in the local RDC calculator from the Forman-Kay group[14]. Because local back-calculation of RDCs has been shown to be able to better model experimental RDCs of disordered states when using smaller ensembles of structures[3], we use the local RDC back-calculator from the Forman-Kay lab[14] to get per-conformation RDCs for the amide bond vector of each residue in the target ensemble. For X-EISD scoring, we estimate the uncertainty in back-calculation error $\sigma_{qRDC} = 0.88$ Hz based on the standard deviation evaluated on the test set of peptides in the local RDC publication.[14] We set $\sigma_{RDCex} = 1.0$ Hz given the experimental data that was deposited in the Protein Ensemble Databank (PED)[41, 42] for the drkN SH3 domain[15].

*Hydrodynamic Radius ($R_h$)*. The hydrodynamic radius can be experimentally determined by calculating the translational diffusion coefficient of the macromolecule with techniques such as pulsed field gradient NMR[15], size exclusion chromatography[43, 44], or dynamic light scattering[45], and then using the Stokes-Einstein relationship to calculate an ensemble-averaged estimate of the $R_h$. We use the program HYDROPRO[46] to calculate $R_h$, which takes static structures and uses a bead-shell model to estimate hydrodynamic properties. For X-EISD scoring, we calculate the ensemble-averaged back-calculated $< R_h >$ over the set of candidate structures, and set the experimental error to $\sigma_{Rhex} = 0.30$ as reported in the original work on the drkN SH3 domain.[15]. Because HYDROPRO is described to have +/-4% error in the estimation of $R_h$, we assign the back-calculation error $\sigma_{qRh} = 0.08$ given the reported experimental value of 20.3 Å.[15]

*Single Molecule Fluorescence Resonance Energy Transfer*. FRET[16, 17, 20] reports on long range distances between two covalently bound dyes through a dipole-dipole non-radiative transfer of energy from the excited-state donor fluorophore to the ground-state acceptor fluorophore. The efficiency of energy transfer, $E$, depends sharply on the on the inter-fluorophore distance, $r_{D-A}$, distance:

$$E = (1 + (r_{D-A}/r_0)^6)^{-1} \tag{9}$$

where $r_0$ is the Förster radius of the donor-acceptor pair. For single-molecule FRET (smFRET) measurements on IDPs and unfolded proteins, the distribution of inter-fluorophore distances is sampled much faster than the typical averaging time of the experiment (~1 ms), such that only an average FRET efficiency, $\langle E \rangle$, is observed.[47] The $\langle E \rangle$ therefore restrains the distribution of distances between two labeled residues. Multiple experiments consisting of different FRET constructs—different pairs of dyes, or dyes linked to different sites in the protein sequence—can be used to produce multiple restraints. There is a possibility that, depending on nature of the dye and the labelling site, they interact with the system and perturb its conformational landscape[48-51], as has been seen for PREs[12], but again can be carefully selected to minimize artifacts.

The $\langle E \rangle$ can be back-calculated by taking the distance measurements from static structures, calculating efficiencies, and then averaging together. Often a model is needed to account for the difference between the distance between the two residues to which dyes would be attached, and the distance between the dye centers themselves. The "scaling up" approach has been previously used to account for the FRET tags, and uses a simple polymer model to scale up the Cα-Cα distance of the native protein[52-54]:

$$r_{D-A} = r_{C\alpha-C\alpha} \left(\frac{N+N_{linker}}{N}\right)^v \tag{10}$$

where $r_{C\alpha-C\alpha}$ is the Cα-Cα distance, $N$ is the number of residues between the relevant residues, $N_{linker}$ is the number of estimated additional amino acids, and $v$ is the Flory scaling exponent. To estimate the back-calculation uncertainty $\sigma_{qFRET}$, we calculate the variation in back-calculated FRET efficiency that results from varying the parameters $N_{linker}$, $v$, and $r_0$ as discussed by Gomes and co-workers[47] and further described in Figure S2. We arrive at a value of $\sigma_{qFRET} = 0.006$, and we use a typical estimate of the experimental uncertainty of 0.02 for $\sigma_{FRETex}$.

*Small Angle X-ray Scattering (SAXS).* SAXS has been a powerful tool for categorization of IDPs in their monomeric state as collapsed semi-ordered ensembles, collapsed disordered ensembles, or extended disordered ensembles.[55-58] The most well-known back-calculator from structure to SAXS intensity curves is the CRYSOL software program[18], and for all members of the ensemble we calculate an intensity curve and then average to obtain the SAXS observable. For X-EISD we treat each intensity point as an independent measurement and scored according to the simple X-EISD formulation like individual chemical shifts via Eq. (5). The back-calculation uncertainty $\sigma_{qSAXS} = 0.006$ is estimated by calculating overall RMSDs of the intensity points along the curve for a set of optimized ensembles. We use the experimental uncertainty estimate $\sigma_{SAXSex} = 0.008 - 0.02$, with larger uncertainty near Q = 0 and decreasing toward larger values of Q. We follow the protocol laid out by Sedlak and co-workers to quantify measurement errors incurred in SAXS experiments.[19]

In order to test the X-EISD Bayesian approach for these various data types, we consider the unfolded state of the drkN SH3 domain.[15, 20, 21] The drkN SH3 domain is in slow exchange on the NMR timescale between folded and unfolded states under typical buffer conditions that are not either denaturing or stabilizing, and in this work we only consider the unfolded state. For the chemical shift, J coupling, NOE, PRE, and RDC data, because of the distinct signals for the unfolded and folded states of the drkN SH3 domain, we directly use only the unfolded state NMR data. For $R_h$ and SAXS, we use the procedure applied by Forman-Kay and co-workers previously[15] of taking the measured experimental data for the exchanging equilibrium state, the experimental data for the stabilized folded state, and the known fraction of the folded state present at equilibrium and subtracting out the effect of the folded state to obtain experimental data for just the unfolded state of the peptide. For smFRET, we ignore the peak at <E> = 1.0, representing the folded state, and score and optimize only using the peak at 0.55, assuming that this population represents the unfolded conformations. The total data set includes 267 chemical shifts, 47 J-couplings, 93 homonuclear NOE distance restraints, 68 PRE distance restraints, 28 RDCs, a SAXS intensity curve, hydrodynamic radius, $R_h$, and smFRET efficiency data[20].

We rank and optimize three different starting pools of structures for the unfolded state of the drkN SH3 domain. The first is a collection of ~100,000 conformations consisting of a random coil ensemble generated with the TraDES program[31], and which is unoptimized with respect to the experimental data (called RANDOM). We also consider an optimized ensemble generated with the ENSEMBLE program that is comprised of 1,700 conformations and is available through the PED[41, 42], and which was generated using all of the same NMR data types except for the smFRET efficiency data (called ENSEMBLE). Figure 1 shows that the underlying structural picture is quite different between the RANDOM and ENSEMBLE starting pool of structures, such as the percentage of secondary structure type for each residue averaged over the pool, and global characteristics embodied in the distribution of the radius of gyration. In particular the ENSEMBLE pool is characterized by high helix propensity and small amounts of parallel-beta sheet for residues 16-20, and some helical content over residues 30-45, unlike the featureless RANDOM ensemble dominated by bends and turns but no population of helical or β–sheet structure. The RANDOM starting pools exhibits a bimodal $R_g$ distribution with $<R_g>$ of 21.2 ± 0.8 Å, whereas the ENSEMBLE shows a very tight unimodal distribution of $<R_g>$ of 18.5 ± 0.3 Å.

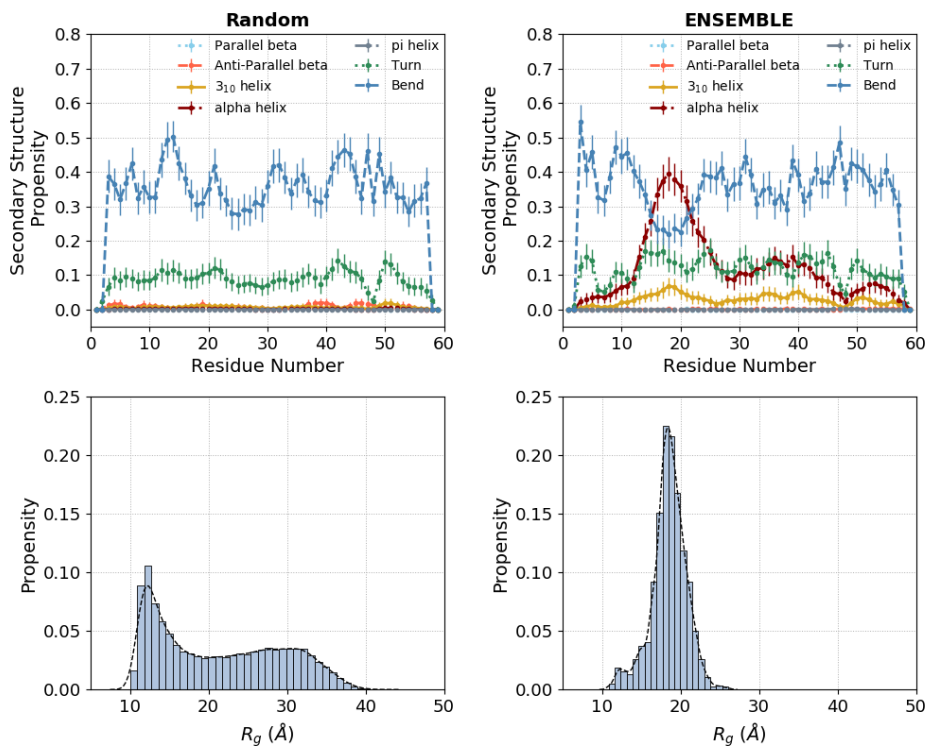

**Figure 1**: *Secondary structure propensities per residue and radius of gyration of the drkN Sh3 domain for the unoptimized RANDOM (left) and ENSEMBLE (right) starting pools.* Error bars are shown as ± one standard deviation for the secondary structure propensities of 1,000 random sampling ensembles of 100 conformers each from the two starting structural pools with no X-EISD score optimization applied.

Table 1 provides the X-EISD scores and RMSD error per experimental data type for the unoptimized RANDOM and ENSEMBLE starting pools of structures (see online Methods). Having already been refined against the full set of experimental data, the ENSEMBLE starting pool is an obviously better ensemble when compared the initial RANDOM ensemble by X-EISD score for RDCs, J-couplings, and chemical shifts. However, although largely equivalent in regards SAXS score, the other data types are inconclusive given the large standard deviation (STDs) in the pool. Furthermore, the experimental and back calculations errors ($\sigma_{exp}$ and $\sigma_q$, respectively, see Methods) are much smaller than the STDs, indicating that we can define an ensemble with higher probability than the original RANDOM and ENSEMBLE structural pools. Hence Table 1 also provides the RANDOM and ENSEMBLE scores after a basic MCMC optimization using the X-EISD probability function for all of the experimental data and data types. This is not meant to be an exhaustive optimization, but just to show if the experimental and back-calculation uncertainties can permit further optimization of the two ensembles.

**Table 1**: *X-EISD scores and RMSDs for all experimental data types for unoptimized and optimized ensembles for the drkN SH3 domain unfolded state using* RANDOM, ENSEMBLE, *and* MIXED *starting pools. The values in parentheses are the standard deviations that reflect variations among the 1,000 independent repeats of sub-ensembles of 100 conformations each, before and after optimization.*

| Experimental data type | UNOPTIMIZED | | | |
|---|---|---|---|---|
| | RANDOM | | ENSEMBLE | |
| | X-EISD Score | RMSD | X-EISD Score | RMSD |
| 267 CSs (ppm) | 202.0 (4.4) | 0.58 (0.01) | 263.5 (4.4) | 0.42 (0.008) |
| 47 JCs (Hz) | –82.2 (4.1) | 0.91 (0.01) | 34.4 (1.8) | 0.30 (0.02) |
| 28 RDCs (Hz) | –59.7 (1.1) | 1.22 (0.05) | -51.8 (0.7) | 0.70 (0.05) |
| 93 NOEs (Å) | 497.3 (5.4) | 4.62 (0.24) | 517.7 (5.5) | 3.80 (0.35) |
| 68 PREs (Å) | –234.3 (186.3) | 6.06 (0.72) | 225.2 (191.2) | 3.44 (0.94) |
| smFRET $<E>$ | –18.2 (13.9) | 0.14 (0.04) | 0.35 (5.0) | 0.07 (0.03) |
| $R_h$ (Å) | –0.9 (0.3) | 0.79 (0.30) | -0.37 (0.0) | 0.09 (0.07) |
| SAXS (Intensity) | 372.8 (0.4) | 0.005 (0.001) | 372.8 (0.3) | 0.004 (0.001) |
| Experimental data type | OPTIMIZED | | | |
| | RANDOM | | ENSEMBLE | |
| | X-EISD Score | RMSD | X-EISD Score | RMSD |
| 267 CSs (ppm) | 251.5 (2.7) | 0.51 (0.00) | 270.0 (1.3) | 0.47 (0.00) |
| 47 JCs (Hz) | –33.1 (2.6) | 0.73 (0.01) | 39.7 (0.7) | 0.24 (0.01) |
| 28 RDCs (Hz) | –55.6 (0.6) | 1.00 (0.04) | -50.7 (0.3) | 0.60 (0.03) |
| 93 NOEs (Å) | 526.7 (1.7) | 3.17 (0.11) | 535.8 (0.8) | 2.59 (0.06) |
| 68 PREs (Å) | 451.5 (4.6) | 1.50 (0.19) | 460.4 (4.7) | 1.09 (0.12) |
| smFRET $<E>$ | 7.1 (0.1) | 0.01 (0.01) | 7.1 (0.1) | 0.005 (0.004) |
| $R_h$ (Å) | –0.5 (0.0) | 0.16 (0.11) | -0.6 (0.0) | 0.44 (0.05) |
| SAXS (Intensity) | 373.6 (0.1) | 0.003 (0.000) | 373.5 (0.1) | 0.003 (0.000) |
| Experimental data type | MIXED (unoptimized) | | MIXED (optimized) | |
| | X-EISD Score | RMSD | X-EISD Score | RMSD |
| 267 CSs (ppm) | 249.1 (4.7) | 0.49 (0.01) | 285.0 (1.7) | 0.44 (0.00) |
| 47 JCs (Hz) | -5.9 (6.9) | 0.60 (0.04) | 33.1 (1.1) | 0.32 (0.01) |
| 28 RDCs (Hz) | -54.7 (0.9) | 0.93 (0.06) | -51.0 (0.3) | 0.62 (0.03) |
| 93 NOEs (Å) | 514.0 (4.8) | 3.92 (0.25) | 536.8 (0.9) | 2.50 (0.08) |
| 68 PREs (Å) | 156.6 (156.2) | 3.94 (0.81) | 460.4 (4.6) | 1.28 (0.21) |
| smFRET $<E>$ | -4.7 (8.4) | 0.10 (0.04) | 7.1 (0.0) | 0.004 (0.003) |
| $R_h$ (Å) | -0.5 (0.1) | 0.24 (0.18) | -0.6 (0.1) | 0.44 (0.09) |
| SAXS (Intensity) | 373.2 (0.4) | 0.004 (0.001) | 373.8 (0.1) | 0.002 (0.000) |

Figure 2 shows the difference in structural outcome from the MCMC optimization, which stems from the discriminative ability of the X-EISD formalism to prioritize data types for which we have higher certainty in the experimental and back-calculated data, by "rewarding" that data type over a data type for which there may be more significant experimental or back-calculation uncertainty. We find that the SAXS, RDCs, and $R_h$ data types have not played a significant role in differentiating among the ensembles after MCMC optimization, showing that the most discriminatory power comes from the improvement in NOEs, smFRET, J-coupling, and chemical shifts. Although there is also a big improvement in the PREs, this may be due to our assumption of near-zero back-calculation uncertainty in combination with the relatively small assigned experimental uncertainties, thereby assigning high confidence to this X-EISD module, and producing strong changes in score.

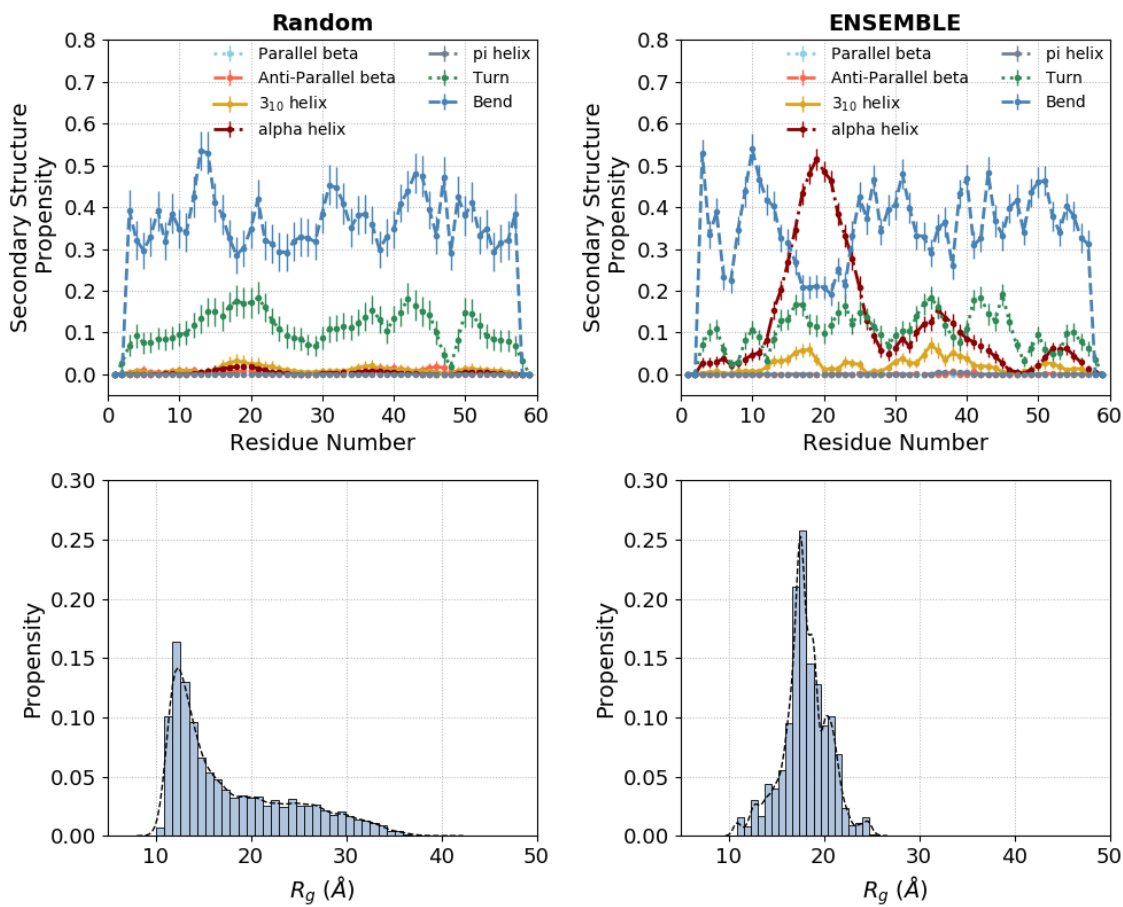

**Figure 2**: *Secondary structure propensities per residue and radius of gyration of the drkN SH3 domain unfolded state for the optimized RANDOM (left) and ENSEMBLE (right) ensembles.* Error bars are shown as ± one standard deviation for the secondary structure propensities of 1,000 random sampling ensembles of 100 conformers each from each of the optimized ensembles.

Even so, the optimized RANDOM ensemble has improved significantly, and performs better than the original unoptimized RANDOM pool or even the original ENSEMBLE data as measured by global characteristics of the chains, i.e. NOEs and smFRET efficiency which shows greater compaction in the $R_g$ distribution with $<R_g>$ of 17.9 ± 0.3 Å (Figure 2). However, it is more poorly scoring in regards local structure relative to the optimized ENSEMBLE, as measured in particular by the J-coupling score and to a lesser extent for the chemical shifts. The optimized ENSEMBLE is better than the original ENSEMBLE with respect to all global and local data type X-EISD scores, and has a secondary assignment that favors greater amounts of helical structure for residues 16-20, 30-45, and 50-55 and an $<R_g>$ of 18.0 ± 0.1 Å.

Although the final optimized ENSEMBLE score indicates that it is a better fit to the data than the optimized RANDOM ensemble, we next consider how sensitive this result is to the available conformers in the selection pool. We created a MIXED starting pool, comprised of 50% from the optimized RANDOM pool and 50% from the optimized ENSEMBLE pool, and Table 1 shows that the score of this unoptimized pool, is largely inferior to the two optimized parent ensembles. However, after a basic MCMC optimization with the X-EISD scoring function, the MIXED pool shifts its composition to 24% RANDOM and 76% ENSEMBLE conformers, with better chemical shift scores that counteract the small deterioration in J-coupling scores relative to the optimized ENSEMBLE parent.

What emerges from the optimization is a structural picture of an ensemble with largely the same local secondary structure features as the ENSEMBLE parent, but a marked decrease in the percentage of α–helix in and around residue 20, and difference in global characteristics with a less compact and broader radius of gyration distribution (with $<R_g>$ = 20.1 ± 0.4 Å) reflecting support for the RANDOM parent conformers (Figure 3). This difference in optimized structural conformational pools between MIXED and ENSEMBLE arises from the balance among the relative changes allowed for the chemical shifts, J-couplings, and NOEs, given their mix of experimental and back-calculation uncertainties. While smFRET scores can be used to optimize an ensemble (the smFRET score typically goes from negative values to 7.1 upon optimization) - it ultimately does not provide discrimination between the two optimized ensembles since they all have a perfect fit. In essence, the MIXED optimized ensemble is as probable as the optimized ENSEMBLE result, but with different sub-populations of structural conformers. This provides an excellent example in which data and data processing uncertainties processed under a Bayesian formalism can yield an alternative structural hypotheses that can stimulate further experiments, unlike methods that indiscriminately fit all of the experimental data.

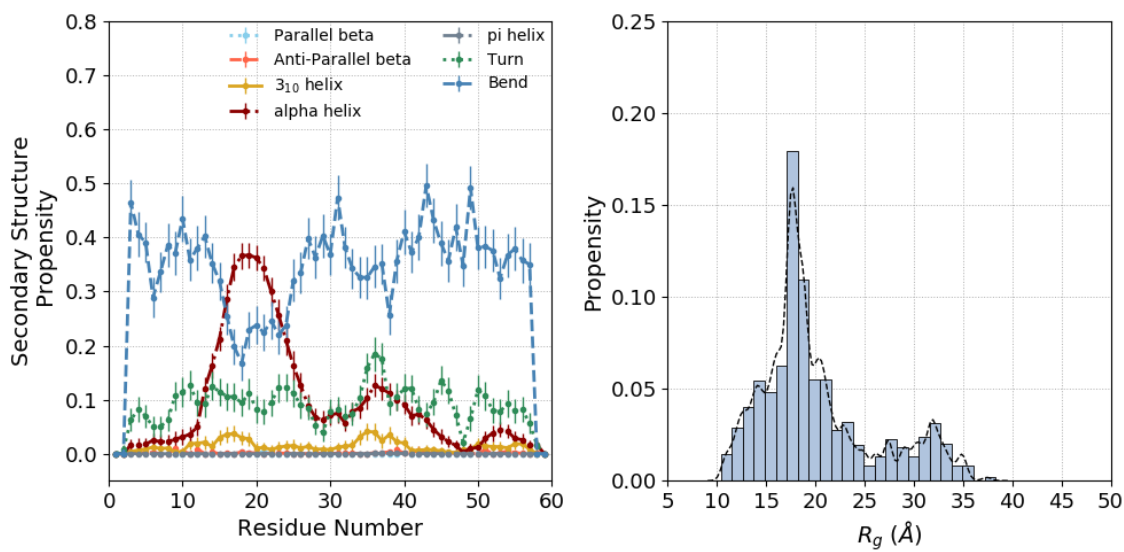

**Figure 3**: *Secondary structure propensities per residue and radius of gyration of the drkN SH3 domain unfolded state for the optimized MIXED ensemble.* Error bars are shown as ± one standard deviation for the secondary structure propensities among the 1,000 independently drawn and optimized ensembles of 100 structures each.

The X-EISD method can also provide guidance as to which experimental data type is most valuable. To show this we run the X-EISD optimization using just a single data type to define the upper bound of the highest score achievable for that data category, and illustrated on the unoptimized RANDOM starting pool. Table 2 shows that optimization against one data type (the diagonal entries) can influence the scores of the unoptimized data types (off-diagonal entries). Although both RDCs and SAXS have multiple restraints, the RDCs have low and lower variability in scores, whereas the SAXS intensity has high scores and low variability in scores, such that each is barely influenced by any single data type optimization. For RDCs there is large uncertainty in the interpretation of the experimental data and back-calculation such that ensembles can be highly variable with no penalty, while for SAXS there is little uncertainty and therefore little score movement. The $R_h$ score can be improved under certain data type optimizations, but with one restraint and fairly large experimental uncertainty, the improvement is marginal. The remaining experimental data types offer interesting mutual support or discord. The FRET score is indirectly optimized through direct optimization of chemical shifts, J-coupling, PREs, and Rh, but worsens under the optimization with the other remaining data types. The J-coupling is most compatible with the chemical shift optimization, NOEs are most compatible with J-coupling optimization, and PREs most benefit from NOE optimization. Although SAXS optimized ensembles slightly improves most data types to some extent, its optimization leads to very poor smFRET and PRE scores. In summary, the single mode optimization would consider that chemical shifts, J-couplings, NOEs, PREs, and FRET can strongly influence, or be strongly influenced, by other data types, and provide the most scope for improving IDP ensemble calculations to obtain agreement with all of the solution data. A similar conclusion was reached in recent work by Gome and co-workers that FRET and PRE provide strong influence on IDP ensemble calculations.[59]

**Table 2:** *Single data type optimization operating on the unoptimized RANDOM ensemble.* Mean scores for all data types, as indicated by the column, resulting from optimizing 1,000 ensembles of 100 structures each by seeking to maximize the X-EISD score of the data type indicated by the row. Numbers in parentheses are standard deviations in score among the 1,000 independently optimized ensembles of each type. The experimental and back calculations errors for NOEs ($\sigma_{exp}$=2.0-2.5; $\sigma_q$=0.0001), PREs ($\sigma_{exp}$=1.3-25.4; $\sigma_q$=0.0001), SAXS ($\sigma_{exp}$=0.008-0.02; $\sigma_q$=0.006), RDCs ($\sigma_{exp}$=1.0; $\sigma_q$=0.9), $R_h$ ($\sigma_{exp}$=0.3; $\sigma_q$=0.08), smFRET <E> ($\sigma_{exp}$=0.02; $\sigma_q$=0.006), CSs ($\sigma_{exp}$=0.03-0.3; $\sigma_q$=0.1-0.5). For JCs ($\sigma_{exp}$=0.5; $\sigma_q$=*), the back-calculation error is handled with individual error parameters for the A, B, C of the Karplus equation.

| Data Type (Optimization) | Data Type (EISD Score) | | | | | | | |
|---|---|---|---|---|---|---|---|---|
| | CS | JC | RDC | NOE | PRE | sm<E> | $R_h$ | SAXS |
| CS | 287.3 (1.2) | -66.8 (3.4) | -57.0 (0.8) | 501.1 (5.4) | -201.2 (164.9) | -7.4 (9.5) | -0.6 (0.2) | 373.4 (0.2) |
| JC | 205.8 (4.3) | 5.7 (-1.2) | -58.0 (0.8) | 511.9 (3.6) | -89.3 (101.9) | -4.1 (8.5) | -0.9 (0.3) | 372.6 (0.4) |
| RDC | 210.5 (4.1) | -79.6 (3.9) | -48.0 (0.0) | 489.9 (6.8) | -406.4 (251.1) | -54.0 (19.6) | -1.8 (0.5) | 371.7 (0.6) |
| NOE | 201.4 (4.7) | -71.7 (4.1) | -57.6 (0.7) | 544.7 (0.5) | -50.0 (102.3) | -18.0 (11.3) | -1.6 (0.4) | 371.8 (0.5) |
| PRE | 203.9 (4.4) | -79.9 (4.2) | -59.0 (1.0) | 501.7 (4.9) | 465.1 (3.8) | 2.0 (-6.1) | -0.5 (0.1) | 373.5 (0.2) |
| sm<E> | 200.3 (4.5) | -79.2 (4.2) | -58.8 (0.9) | 505.3 (4.0) | -101.1 (135.0) | 7.2 (0.0) | -0.5 (0.0) | 373.4 (0.2) |
| $R_h$ | 200.9 (4.4) | -80.2 (4.3) | -59.1 (1.0) | 503.4 (4.3) | -145.8 (149.8) | 2.9 (-3.4) | -0.4 (0.0) | 373.3 (0.2) |
| SAXS | 207.1 (4.1) | -78.0 (4.0) | -57.4 (0.7) | 487.1 (6.4) | -1094.0 (592.4) | -41.7 (9.5) | -0.4 (0.0) | 374.2 (0.0) |
| Random unoptimized | 202.0 (4.4) | -82.2 (4.1) | -59.7 (1.1) | 497.3 (5.4) | -234.3 (186.3) | -18.2 (13.9) | -0.9 (0.3) | 372.8 (0.4) |

**Table 3:** *Dual X-EISD optimized scores using SAXS with one other data type operating on the unoptimized RANDOM pool and resulting $<R_g>$ values for single and dual optimization.* Experimental and back calculations uncertainties given in Table 2.

### EISD Score

| Second Data Type (Optimization) \ Data Type (EISD Score) | CS | JC | RDC | NOE | PRE | sm<E> | $R_h$ | SAXS |
|---|---|---|---|---|---|---|---|---|
| CS | 287.2 (1.2) | -66.1 (3.3) | -56.8 (0.8) | 502.7 (4.7) | -170.6 (156.2) | -0.2 (5.3) | -0.5 (0.1) | 373.7 (0.1) |
| JC | 206.9 (4.2) | 5.4 (1.2) | -58.0 (0.8) | 509.4 (3.8) | -111.1 (112.1) | 5.6 (1.9) | -0.5 (0.1) | 373.4 (0.2) |
| RDC | 211.5 (4.0) | -76.5 (3.9) | -48.1 (0.0) | 499.7 (4.6) | -189.9 (167.8) | 1.5 (3.5) | -0.5 (0.0) | 374.0 (0.0) |
| NOE | 205.2 (4.5) | -73.7 (4.0) | -57.6 (0.8) | 544.2 (0.5) | -140.1 (133.1) | 6.5 (0.9) | -0.5 (0.1) | 373.7 (0.1) |
| PRE | 204.5 (4.5) | -78.6 (4.0) | -58.2 (0.8) | 502.0 (4.6) | 465.3 (3.3) | 4.0 (3.3) | -0.5 (0.1) | 373.9 (0.1) |
| sm<E> | 207.2 (4.2) | -78.5 (4.2) | -57.6 (0.7) | 491.0 (6.4) | -252.3 (229.6) | 7.2 (0.0) | -0.4 (0.0) | 374.2 (0.0) |
| $R_h$ | 207.3 (4.2) | -78.1 (4.1) | -57.3 (0.7) | 489.0 (6.7) | -1076.0 (546.2) | -39.0 (9.4) | -0.4 (0.0) | 374.2 (0.0) |
| SAXS | 207.1 (4.1) | -78.0 (4.0) | -57.4 (0.7) | 487.1 (6.4) | -1094.0 (592.4) | -41.7 (9.5) | -0.4 (0.0) | 374.2 (0.0) |
| Random unoptimized | 202.0 (4.4) | -82.2 (4.1) | -59.7 (1.1) | 497.3 (5.4) | -234.3 (186.3) | -18.2 (13.9) | -0.9 (0.3) | 372.8 (0.4) |
| Random optimized | 251.5 (2.7) | -33.1 (2.6) | -55.6 (0.6) | 526.7 (1.7) | 451.5 (4.6) | 7.1 (0.1) | -0.5 (0.0) | 373.6 (0.1) |

### Average Radius of Gyration ($<R_g>$)

| Second Data Type \ First Data Type | CS | JC | RDC | NOE | PRE | sm<E> | $R_h$ | SAXS |
|---|---|---|---|---|---|---|---|---|
| CS | 20.0 (0.7) | 18.4 (0.3) | 19.8 (0.6) | 17.9 (0.6) | 19.3 (0.6) | 18.1 (0.6) | 19.5 (0.5) | 19.2 (0.4) |
| JC |  | 18.6 (0.4) | 19.3 (0.3) | 17.6 (0.3) | 19.3 (0.4) | 17.9 (0.3) | 18.8 (0.3) | 17.8 (0.1) |
| RDC |  |  | 22.9 (0.5) | 17.6 (0.6) | 19.6 (0.7) | 16.7 (0.6) | 20.2 (0.4) | 18.6 (0.2) |
| NOE |  |  |  | 15.6 (0.5) | 16.7 (0.5) | 15.4 (0.5) | 17.6 (0.4) | 17.6 (0.3) |
| PRE |  |  |  |  | 18.6 (0.9) | 18.1 (0.6) | 18.2 (0.6) | 17.7 (0.3) |
| sm<E> |  |  |  |  |  | 16.6 (0.6) | 17.3 (0.5) | 17.5 (0.4) |
| $R_h$ |  |  |  |  |  |  | 19.5 (0.4) | 18.2 (0.1) |
| SAXS |  |  |  |  |  |  |  | 18.2 (0.1) |

Table 3 (and Tables S2-S8 in online Methods) provide the results for a dual optimization procedure for which two data types are jointly optimized, such that their simultaneous optimization can influence the scores of the remaining data types which have not contributed to the optimization. The importance of SAXS is its joint optimization always improves the X-EISD scores for other data types, while joint optimization of <E> with J-couplings, NOEs, PREs, and even RDCs goes the farthest in reaching the finalized optimized smFRET score. The consequences are manifest in the independent assessment of $<R_g>$ under the single and joint optimization schemes (Table 3). It is evident that smFRET and NOEs contribute to more collapsed ensembles whereas chemical shifts, RDCs and $R_h$ contribute to more expanded ensembles on average. The expanded CS ensemble is shifted to smaller $<R_g>$ by J-Couplings and smFRET, while SAXS strongly influences all data types. In fact SAXS contributes to the most consistent ensemble average, and provides a corrective measure when jointly optimization with all other data types, yielding something close to the final optimized RANDOM $<R_g>$=17.9 ± 0.3 Å with the sole exception of chemical shifts.

**DISCUSSION**

We have developed a Bayesian scoring formalism for a large variety of solution experimental data types, spanning those that report on very local to very global structural information. The X-EISD approach is able to account for varying levels of uncertainty in both experiment and back-calculation for each data type, making it distinct from other Bayesian approaches, while the very good O(N) scaling with ensemble size facilitates the high number of replicates we can perform, demonstrating the cost-effectiveness of the algorithm. In the future[59], the X-EISD scoring can be utilized within more sophisticated optimization approaches, as well as operating on Boltzmann weighted ensembles derived from state-of-the-art force fields and sampling methods.[3, 12, 60-62] One of the primary results we have demonstrated is that certain experimental data types provide more value than others for influencing the most probable disordered state ensemble, which can only be understood through a Bayesian formalism that recognizes their differences. Because of this, we have shown that two equally probable disordered state ensembles are both consistent with experimental and back-calculation uncertainties for the drk SH3 unfolded state domain, generating new hypotheses about function given their differences in weighting of sub-populations of conformational states.

**ACKNOWLEDGEMENT**. We thank the National Institutes of Health for support under Grant 5R01GM127627-03. J.D.F.-K. also acknowledges support from the Natural Sciences and

Engineering Research Council of Canada (NSERC) grant RGPIN-2016-06718 and the Canada Research Chairs program. C.C.G. thanks NSERC a for support under RGPIN 2017–06030. C.C.G. thanks the Natural Sciences and Engineering Research Council of Canada for support under RGPIN 2017–06030. This research used the computational resources of the National Energy Research Scientific Computing Center, a DOE Office of Science User Facility supported by the Office of Science of the U.S. Department of Energy under Contract No. DE-AC02-05CH11231.

**Conflict of interest.** The authors declare that they have no conflict of interest.

**Author contribution.** THG, J.L. and J.D.F.-K. conceived the scientific content and direction; J.L. and M.H. performed the calculations; J.D.F-K., M.K., G.N.G and C.C. provided experimental data and analysis; J.L. and THG wrote the manuscript; J.L. and M.H. created the Figures. All authors contributed insights and discussed and edited the manuscript.